\documentclass[useAMS,usenatbib]{mn2e}

\usepackage{aas_macros}  
\usepackage{bm}          
\usepackage{graphicx}    
\usepackage{amsmath}
\usepackage{etoolbox}
\usepackage[colorlinks=true, urlcolor={blue},linkcolor={black},citecolor={black},pdfborder={0 0 0}]{hyperref}

\usepackage{tikz}        
\usetikzlibrary{calc}    
\usepackage{printlen}

\newcommand{\posterior}{\mathcal{P}}

\newcommand{\lik}{\mathcal{L}}
\newcommand{\prior}{\pi}
\newcommand{\ev}{\mathcal{Z}}
\newcommand{\data}{\mathcal{D}}
\newcommand{\params}{\theta}

\newcommand{\Prob}{\text{P}}
\newcommand{\model}{\mathcal{M}}

\newcommand{\PolyChord}{{\sc PolyChord}}
\newcommand{\MultiNest}{{\sc MultiNest}}
\newcommand{\FORTRAN}{{\sc FORTRAN}}
\newcommand{\openMPI}{{\sc openMPI}}
\newcommand{\CosmoMC}{{\sc CosmoMC}}
\newcommand{\CAMB}{{\sc CAMB}}
\newcommand{\CosmoChord}{{\sc CosmoChord}}

\newcommand{\nlive}{n_\mathrm{live}}
\newcommand{\ndims}{n_\mathrm{dims}}
\newcommand{\nprocs}{n_\mathrm{procs}}
\newcommand{\nlike}{N_\mathcal{L}}

\title[\PolyChord]{\PolyChord{}: nested sampling for cosmology}
\author[W.J.~Handley et.\ al]
{ W.J.~Handley$^{1,2}$\thanks{wh260@mrao.cam.ac.uk}, M.P.~Hobson$^1$\thanks{mph@mrao.cam.ac.uk} \& A.N.~Lasenby$^{1,2}$\thanks{a.n.lasenby@mrao.cam.ac.uk} \\
$^1$Astrophysics Group, Cavendish Laboratory, J.~J.~Thomson Avenue, Cambridge, CB3 0HE, UK \\
$^2$Kavli Institute for Cosmology, Cambridge, Madingley Road, Cambridge, CB3 0HA, UK
}

\date{Received 6 February 2015}
\pubyear{2015}

\begin{document}


\makeatletter
\patchcmd{\NAT@test}{\else \NAT@nm}{\else \NAT@nmfmt{\NAT@nm}}{}{}

\DeclareRobustCommand\citepos
  {\begingroup
   \let\NAT@nmfmt\NAT@posfmt
   \NAT@swafalse\let\NAT@ctype\z@\NAT@partrue
   \@ifstar{\NAT@fulltrue\NAT@citetp}{\NAT@fullfalse\NAT@citetp}}

\let\NAT@orig@nmfmt\NAT@nmfmt
\def\NAT@posfmt#1{\NAT@orig@nmfmt{#1's}}

\pagerange{\pageref{firstpage}--\pageref{lastpage}} 
\pubyear{2014}

\maketitle

\label{firstpage}

\begin{abstract}
  \PolyChord{} is a novel nested sampling algorithm tailored for high dimensional parameter spaces. In addition, it can fully exploit a hierarchy of parameter speeds such as is found in \CosmoMC{} and \CAMB{}. It utilises slice sampling at each iteration to sample within the hard likelihood constraint of nested sampling. It can identify and evolve separate modes of a posterior semi-independently and is parallelised using \openMPI{}. \PolyChord{} is available for download at: \url{http://ccpforge.cse.rl.ac.uk/gf/project/polychord/}
\end{abstract}

\section{Introduction}

Over the past two decades, the quantity and quality of astrophysical
and cosmological data has increased substantially.  In response to
this, Bayesian methods have been increasingly adopted as the standard
inference procedure.

Bayesian inference consists of {\em
  parameter estimation\/} and {\em model comparison}.  Parameter
estimation is generally performed using Markov-Chain Monte-Carlo
(MCMC) methods, such as the Metropolis Hastings (MH) algorithm and its
variants \citep{Mackay}.  In order to perform model comparison, one
must calculate the {\em evidence\/}; a high-dimensional integration of
the likelihood over the prior density \citep{Sivia}.  MH methods
cannot compute this on a usable timescale, hindering the use of
Bayesian model comparison in cosmology and astroparticle physics.

A contemporary methodology for computing evidences and posteriors
simultaneously is provided by Nested Sampling
\citep{skilling2006}. This has been successfully implemented in the
now widely adopted 
algorithm \MultiNest\, \citep{MultiNest1,MultiNest2,MultiNest3}.  Modern 
cosmological
likelihoods now involve a large number of parameters, with a hierarchy
of speeds.  \MultiNest\ struggles with high-dimensional parameter
spaces, and is unable to take advantage of this separation of speeds.
\PolyChord{} aim to address these issues, providing a
means to sample high-dimensional spaces across a hierarchy of
parameter speeds.

This letter is a brief overview of our algorithm, which is now in use
in several cosmological applications \citep{planck2015-a24}. It will be followed in the near future by a longer and more
extensive paper. 

\PolyChord{} is available for download from the link at the end of the paper.

\section{Bayesian Inference}

Given some dataset $\data$, one may use a model $\model$ with parameters $\params$ to calculate a likelihood
${\Prob(\params|\data,\model)\equiv\lik_\model(\params)}$. This may be
inverted using Bayes' theorem to find the posterior distribution on
the parameters:
\begin{equation}
  \Prob(\params|\data,\model) 
  \equiv
  \posterior_\model(\params) 
  = 
  \frac{\lik_\model(\params)\prior_\model(\params)}{\ev_\model},
  \label{eqn:Bayes_Abbrv}
\end{equation}
where ${\Prob(\theta|\model)\equiv\prior_\model(\params)}$ is our prior degree of belief on the values of the parameters and ${\Prob(\data|\model)\equiv\ev_\model}$ is the evidence or marginal likelihood, calculated as:
\begin{equation}
  \ev_\model = \int \lik_\model(\params)\prior_\model(\params)d\params.
  \label{eqn:ev_calc}
\end{equation}
One may use the evidence $\ev_\model$ to perform model comparison, comparing $\ev_\model$ to the evidences of other competing models $\{\model_1,\model_2,\ldots\}$.
Applying Bayes theorem to model comparison yields:
\begin{equation}
  \Prob(\model_i|\data) = \frac{\Prob(\data|\model_i)\Prob(\model_i)}{\Prob(\data)} = \frac{\ev_i\pi_i}{\sum_j \ev_j\pi_j},
\end{equation}
where $\Prob(M_i)\equiv\pi_i$ is the prior probability of a model $\model_i$ (usually taken as uniform). Since the evidences $\ev_i$ play a key role in Bayesian model comparison, so their calculation is vital for this aspect of inference.
For a full discussion of Bayesian inference, we recommend \cite{Sivia}
and part IV of \cite{Mackay}.

\section{Nested Sampling}

\cite{skilling2006} constructed an ingenious method of sampling which enables one to estimate the evidence integral~(\ref{eqn:ev_calc}) efficiently. 

One begins by drawing $n$ ``live points'' uniformly from the prior. 
At iteration $i$, the point with the lowest likelihood $\lik_i$ is replaced by a new live point drawn uniformly from the prior with the constraint that its likelihood $\lik>\lik_i$. 
One may estimate the fraction of the prior $X_i$ enclosed by the iso-likelihood contour $\lik_i$, since it will be the lowest volume of $n$ volumes drawn uniformly from $[0,X_{i-1}]$. $X_i$ thus satisfies:
\begin{equation}
  X_i = t_i X_{i-1},\quad P(t_i) = n t_i^{n-1},\quad X_0=1 
  \label{eqn:Xi}
\end{equation}
We may then calculate the evidence by quadrature:
\begin{equation}
  \ev \approx \sum_{i=1}  \lik_i(X_{i-1}-X_i).
  \label{eqn:quadrature}
\end{equation}
and the mean and variance of the random variable $\ev$ may be inferred using equation~(\ref{eqn:Xi}). The posterior mass contained within the live points may be estimated as
${\ev_\mathrm{live} \approx \langle \mathcal{L} \rangle_\mathrm{live} X_\mathrm{live}}$, and convergence is reached when $\ev_\mathrm{live}$ is some small fraction of $\ev$ \citep{keeton}.

In addition, the pool of dead points may be used to construct a set of posterior samples for use in parameter estimation.
For further details on the nested sampling procedure, see \cite{Sivia}.

\section{Sampling within an iso-likelihood contour}
The difficult step in nested sampling is sampling from the prior subject to the hard likelihood constraint $\lik>\lik_i$. The prior may be sampled using inverse transform sampling, so that sampling effectively occurs in the {\em unit hypercube} \citep{MultiNest2}. How one samples in this space under the hard likelihood constraint is where variations of the algorithm differ. 

The \MultiNest\ algorithm samples by using the live points to
construct a set of intersecting ellipsoids which together aim to enclose
the likelihood contour, and then samples by rejection sampling within
it. Whilst being an excellent algorithm for small number of
parameters, any rejection sampling algorithm has an exponential
scaling with dimensionality.

Galilean Sampling \citep{GalileanNestedSampling,Betancourt2011}
samples by an MCMC-chain based approach using reflection off
iso-likelihood contours. This however suffers from the need to tune
parameters, and the requirement of likelihood gradients.

Diffusive nested sampling \citep{DiffusiveNestedSampling} is an alternative and
promising variation on \citepos{skilling2006} algorithm, which utilises MCMC to explore a
mixture of nested probability distributions. Since it is MCMC based, it scales
well with dimensionality. In addition, it can deal with multimodal and
degenerate posteriors, unlike traditional MCMC. It does however have multiple
tuning parameters.

We find that \citepos{NealSlice} slice sampling is well suited to the
problem of drawing points from within an iso-likelihood contour.  In
the one-dimensional case, he suggests the sampling procedure detailed
in Figure~\ref{fig:1d_slice}.  In the $N$-dimensional case, he
suggests multiple methods, the most simple being to perform one
dimensional slice sampling of each parameter in turn (in a random
order). This method has been implemented in nested sampling by
\cite{SystemsBio} to great effect on some simple examples. However,
like Gibbs and MH sampling, this methodology struggles with degenerate
distributions. \PolyChord{} aims to address this issue.

\begin{figure}
  \centerline{%
    \includegraphics{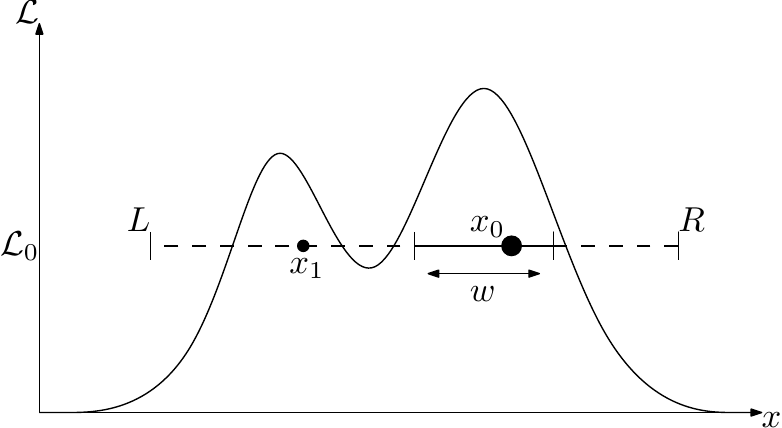}
  }

  \caption{Slice sampling in one dimension. 
    Given a likelihood slice $\lik_0$, a seed point $x_0$ and an initial parameter $w$, slice sampling generates a new point $x_1$ within the horizontal region defined by $\lik>\lik_0$. A point $x$ is within the ``slice'' if $\lik(x)>\lik_0$.
    External bounds are first set $\hat{L}<x_0<\hat{R}$ by expanding a random initial bound of width $w$ until they lie outside the slice via the {\em stepping out\/} procedure. 
    $x_1$ is then sampled uniformly within these bounds.  
    If $x_1$ is not in the slice, then the bound is contracted down to $x_1$, and $x_1$ is then drawn again from these new bounds.
    This procedure is guaranteed to generate a new point $x_1$, and satisfies detailed balance $P(x_0|x_1) = P(x_1|x_0)$. Thus, if $x_0$ is drawn from a uniform distribution within the slice, so is $x_1$. 
    \label{fig:1d_slice}
  }
\end{figure}

\section{The \PolyChord{} Algorithm}

\PolyChord{} implements several novel features compared to \citepos{SystemsBio} slice-based nested sampling. 
We utilise slice sampling in a manner that uses the information in the live points to deal with degeneracies in the contour. 
We also implement a general clustering algorithm which identifies and evolves separate modes in the posterior semi-independently. 
The algorithm is written in \FORTRAN{}95 and parallelised using \openMPI{}. 
It is optimised for the case where the dominant cost is the generation of a new live point. 
This is frequently the case in astrophysical applications, either due to high dimensionality, or to costly likelihood evaluation.
In addition, it has the option of implementing fast-slow parameters, which is extremely effective in its combination with \CosmoMC{}. This is termed \CosmoChord, which may be downloaded from the link at the end of the paper.

Our generic $N$-dimensional slice-sampling procedure is detailed in Figure~\ref{fig:Nd_slice}. 

\subsection{Dealing with degenerate contours}
To solve degenerate cases, we make a linear transformation to ``whiten'' the space, so that in general the contour has size $\sim\mathcal{O}(1)$ in every direction. Uniform sampling is preserved under affine transformations, so this strategy is valid.  

In order to find such a linear transformation, note that  at every step of the algorithm one has some knowledge of the dimensions of the contour, provided by the set of live points and all of the inter-chain points.  
We take the covariance matrix of these points as a reasonable descriptor of the correlation. 
The Cholesky decomposition $T$ of the covariance matrix $\Sigma = T T^{T}$ is a good choice of transformation so that:
\begin{equation}
  T^{-1}{\bf x} =  {\bf y}
  \label{eqn:cholesky}
\end{equation}
transforms from $\bf x$ in the unit hypercube to  $\bf y$ in the {\em sampling space}. 
Working in the sampling space, a reasonable approach that mixes well is to start from a random live point as the seed, choose a randomly oriented orthonormal basis, and sample along these directions in a random order. 
For difficult distributions, one may require multiple iterations of this procedure to ensure that the final point is fully decorrelated from the start point. 
This is illustrated further in Figure~\ref{fig:Nd_slice}.
Within the sampling space, the initial step size may be chosen as $w=1$.

This procedure has the advantage of being dynamically adaptive, and requires no tuning parameters. However, this ``whitening'' process is ineffective for pronounced curving degeneracies.

\begin{figure*}
  \centerline{%
    \includegraphics[width=\textwidth]{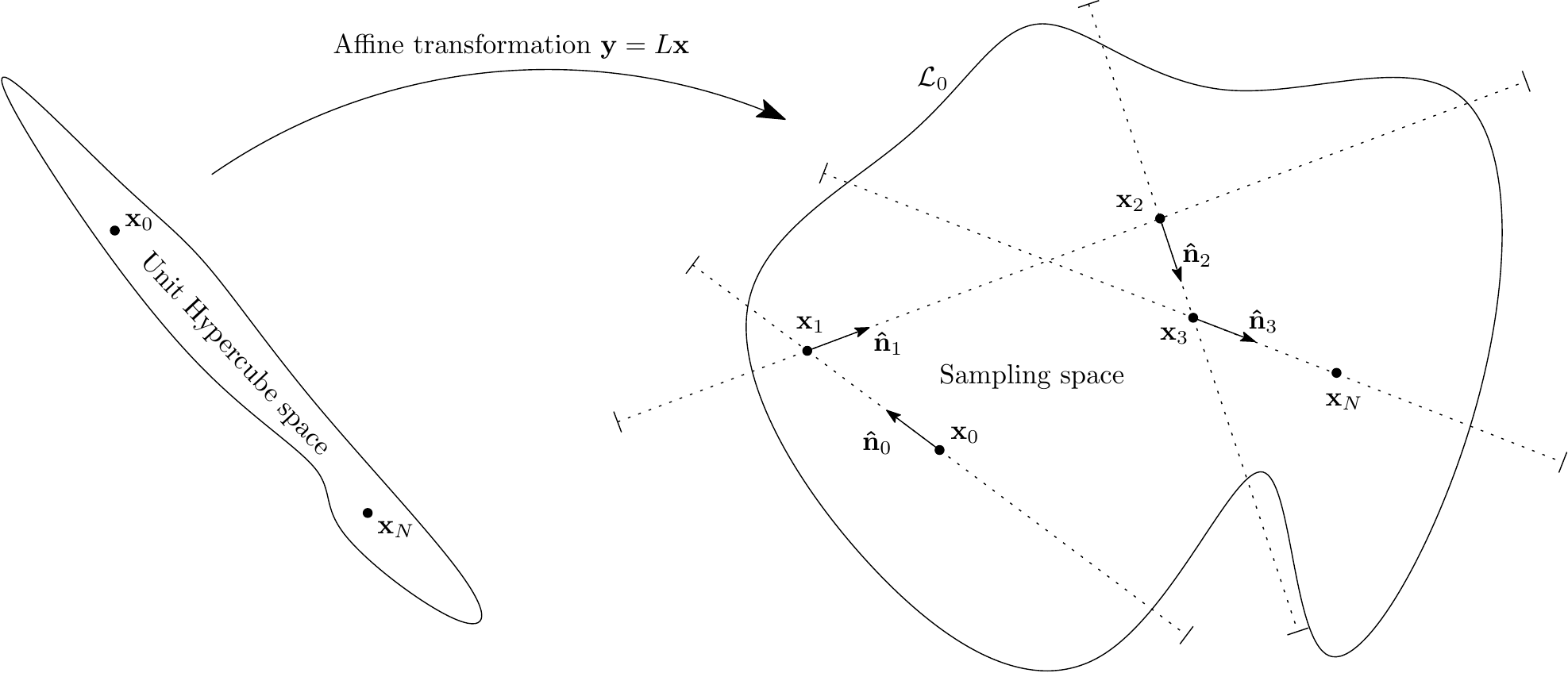}
}
\caption{%
  Slice Sampling in $N$ dimensions. 
  We begin by ``whitening'' the unit hypercube by making a linear transformation which turns a degenerate contour into one with dimensions $\sim\mathcal{O}(1)$ in all directions. 
  This is a linear skew transformation defined by the inverse of the Cholesky decomposition of the live points' covariance matrix. 
  We term this whitened space the {\em sampling space}. 
  Starting from a randomly chosen live point $x_0$, we pick a random direction and perform one dimensional slice sampling in that direction (Figure~\protect\ref{fig:1d_slice}), using $w=1$ in the sampling space. 
  This generates a new point $x_1$ in $\sim\mathcal{O}(\text{a few})$ likelihood evaluations. 
  This process is repeated $\sim\mathcal{O}(\ndims)$ times to generate a new uniformly sampled point $x_N$ which is decorrelated from $x_0$.
\label{fig:Nd_slice}
}
\end{figure*}

\subsection{Clustering}
Multi-modal posteriors are a challenging problem for any sampling algorithm. ``Perfect'' nested sampling (i.e.\ the entire prior volume enclosed by the iso-likelihood contour is sampled uniformly) in theory solves multi-modal problems as easily as uni-modal ones. In practice however, there are two issues.

First, one is limited by the resolution of the live points. If a given mode is not populated by enough live points, it runs the risk of ``dying out''. Indeed, a mode may be entirely missed if the algorithm is not sufficiently resolved. In many cases, this problem can be alleviated by increasing the number of live points.

Second, and more importantly for \PolyChord{}, the sampling procedure may not be appropriate for multi-modal problems. We ``whiten'' the unit hypercube using the covariance matrix of live points. For far-separated modes, the covariance matrix will not approximate the dimensions of the contours, but instead falsely indicate a high degree of correlation.
It is therefore essential for our purposes to have \PolyChord{} recognise and treat modes appropriately.

This methodology splits into two distinct parts;
  (i) recognising that clusters are there, and
  (ii) evolving the clusters semi-independently.

\subsubsection{Recognition of clusters}
Any cluster recognition algorithm can be substituted at this point. 
One must take care that this is not run too often, or one runs the risk of adding a large overhead to the calculation. 
In practice, checking for clustering every $\sim\mathcal{O}(\nlive)$ iterations is sufficient. 
We encourage users of \PolyChord{} to experiment with their own preferred cluster recognition, in addition to that provided and described below. 

It should be noted that the live points of nested sampling are amenable to most cluster recognition algorithms for two reasons.
First, all clusters should have the same density of live points in the unit hypercube, since they are uniformly sampled.
Second, there is no noise (i.e.\ outside of the likelihood contour there will be no live points). Many clustering algorithms struggle when either of these two points are not satisfied.

We therefore choose a relatively simple variant of the $k$-nearest
neighbours algorithm to perform cluster recognition.  If two points are
within one another's $k$-nearest neighbours, then these two points
belong to the same cluster.  We iterate $k$ from $2$ upwards until the
clustering becomes stable (the cluster decomposition does not change
from one $k$ to the next).  If sub-clusters are identified, then this
process is repeated on the new sub-clusters.

\subsubsection{Evolving the clusters semi-independently}
An important novel feature comes from what one does once clusters are identified. 

First, when spawning from an existing live point, the whitening
procedure is now defined by the covariance matrix of the live points
within that cluster. This solves the issue detailed above.

Second, \PolyChord{} would naively spawn live points into a mode with
a probability proportional to the number of live points in that
mode. In fact, what it should be doing is to spawn in proportion to
the volume fraction of that mode. These should be the same, but the
difference between these two ratios will exhibit random-walk like
behaviour, and can lead to biases in evidence calculations, or worse,
cluster death. Instead, one can keep track of an estimate of the
volume in each cluster, using an approached based on (\ref{eqn:Xi}),
and choose the mode to spawn into in proportion to that estimate. This methodology will be fully documented in the more extensive paper.

Thus, the point to be killed off is still the global lowest-likelihood
point, but we control the spawning of the new live point into clusters
by using our estimates of the volumes of each cluster. We call this
`semi-independent', because it retains global information, whilst
still treating the clusters as separate entities.

When spawning within a cluster, we determine the cluster assignment of
the new point by which cluster it is nearest to. It does not matter if
clusters are identified too soon; the evidence calculation will remain
consistent.

\subsection{Parallelisation}
Currently, \PolyChord{} is parallelised by \openMPI{} using a master-slave structure.
One master process takes the job of organising all of the live points, whilst the remaining ${\nprocs-1}$ ``slave'' processes take the job of finding new live points.

When a new live point is required, the master process sends a random live point and the Cholesky decomposition to a waiting slave. 
The slave then, after some work, signals to the master that it is ready and returns a new live point and the inter-chain points to the master.

A point generated from an iso-likelihood contour $\lik_i$ is usable as a new live point for an iso-likelihood contour $\lik_j>\lik_i$, providing it is within both contours. 
One may keep slaves continuously active, and discard any points returned which are not usable. 
The probability of discarding a point is proportional to the volume ratio of the two contours, so if too many slaves are used, then most will be discarded. 
The parallelisation goes as:
\begin{equation}
  \text{Speedup}(\nprocs) = \nlive\log\left[ 1 + \frac{\nprocs}{\nlive} \right],
  \label{}
\end{equation}
and is illustrated in Figure~\ref{fig:parallel}.

\begin{figure}
  \centering
  \includegraphics[width=\columnwidth]{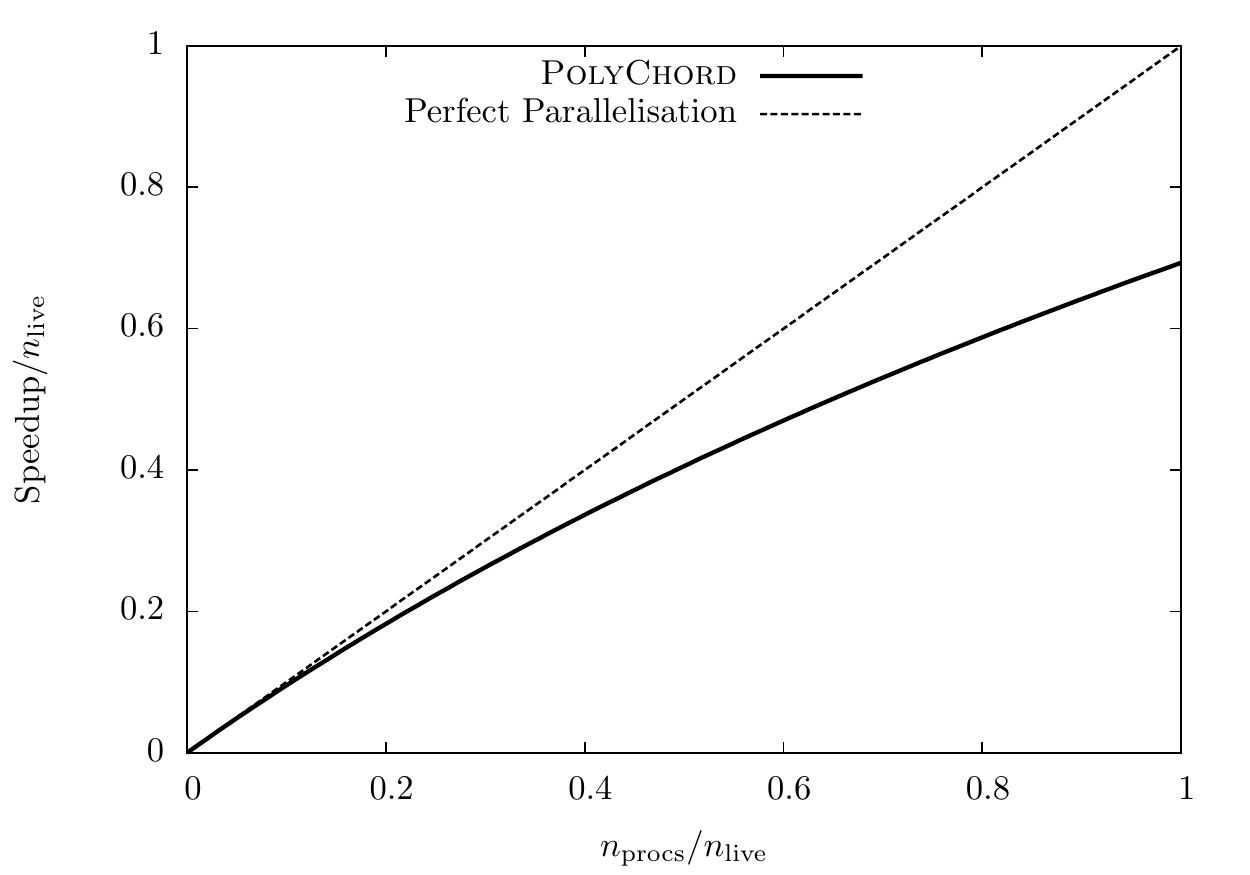}
  \caption{%
Parallelisation of \PolyChord{}. 
The algorithm parallelises nearly linearly, providing that $\nprocs<\nlive$. For most astronomical applications this is more than sufficient.\label{fig:parallel}}
\end{figure}

\subsection{Fast-slow parameters and \CosmoChord}

In cosmological applications, likelihoods exhibit a hierarchy of parameters in terms of calculation speed \citep{LewisFastSlow}. The effect of this is that the likelihoods may be quickly recalculated if one changes certain subsets of the parameters. In \PolyChord{} it is very easy to exploit such a hierarchy. Our transformation to the sampling space is laid out so that if parameters are ordered from slow to fast, then this hierarchy is automatically exploited: A Cholesky decomposition, being a upper-triangular skew transformation, mixes each parameter only with faster parameters.
Further to this, one may use the fast directions to extend the chain length by many orders of magnitude. This helps to ensure an even mixing of live points.

\section{\PolyChord{} in action}
\subsection{Gaussian Likelihood}

As an example of the efficacy of \PolyChord{}, we compare it to
\MultiNest{} on a Gaussian likelihood in $D$ dimensions.  
In both cases, convergence is defined as when the posterior mass contained in the live points is $10^{-2}$ of the total calculated evidence.
We set $\nlive=25D$, so that the evidence error remains constant with $D$. \MultiNest{} was run in its default mode with importance nested sampling and expansion factor $e=0.1$. 
Whilst constant efficiency mode has the potential to
reduce the number of \MultiNest{} evaluations, the low
efficiencies required in order to generate accurate evidences
negate this effect.

With these settings, \PolyChord{} produces consistent evidence and error estimates with an error $\sim0.4$ log units (Figure~\ref{fig:gaussian_evidences}). Using importance nested sampling, \MultiNest{} produces estimates that are within this accuracy.

Figure~\ref{fig:gaussian} shows the number of likelihood
evaluations $\nlike$ required to achieve convergence as a function of dimensionality $D$. 

Even on a
simple likelihood such as this, \PolyChord{} shows a significant
improvement over \MultiNest{} in scaling with dimensionality.
\PolyChord{} at worst scales as ${\nlike\sim\mathcal{O}(D^3)}$, whereas \MultiNest{} has an exponential scaling which emerges in higher dimensions.

However, we must point out that a good rejection algorithm like \MultiNest{} will always win in low dimensions. \MultiNest{} is also extremely effective at navigating pronounced curving degeneracies in low dimensions, whereas \PolyChord{} must use very long chains in order to navigate such contours.

\begin{figure}
  \centering
  \includegraphics[width=\columnwidth]{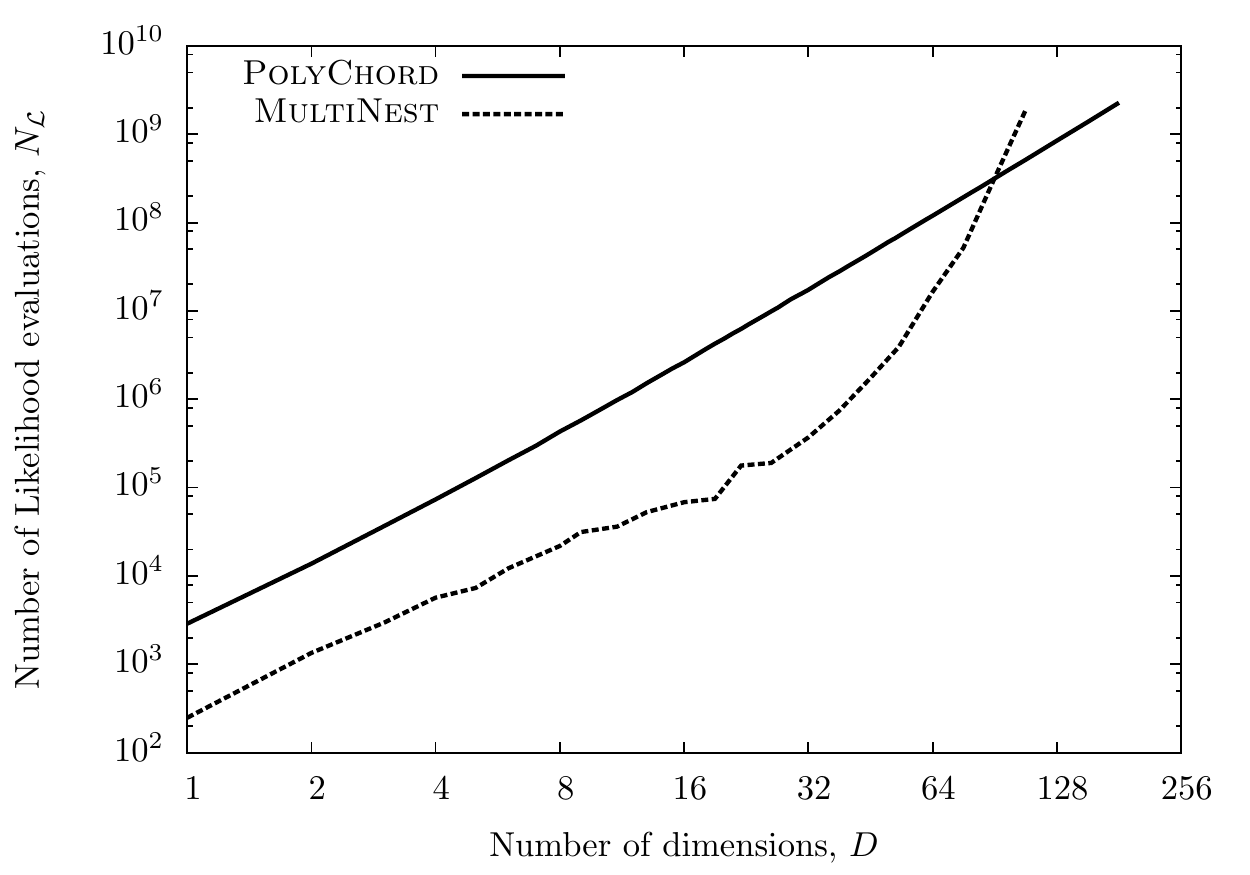}
  \caption{Comparing \PolyChord{} with \MultiNest{} using a
  Gaussian likelihood for different dimensionalities. \PolyChord{} has at worst $\nlike\sim\mathcal{O}(D^3)$, whereas \MultiNest{} has an exponential scaling that emerges at high dimensions.
\label{fig:gaussian}
}
\end{figure}

\begin{figure}
  \centering
  \includegraphics[width=\columnwidth]{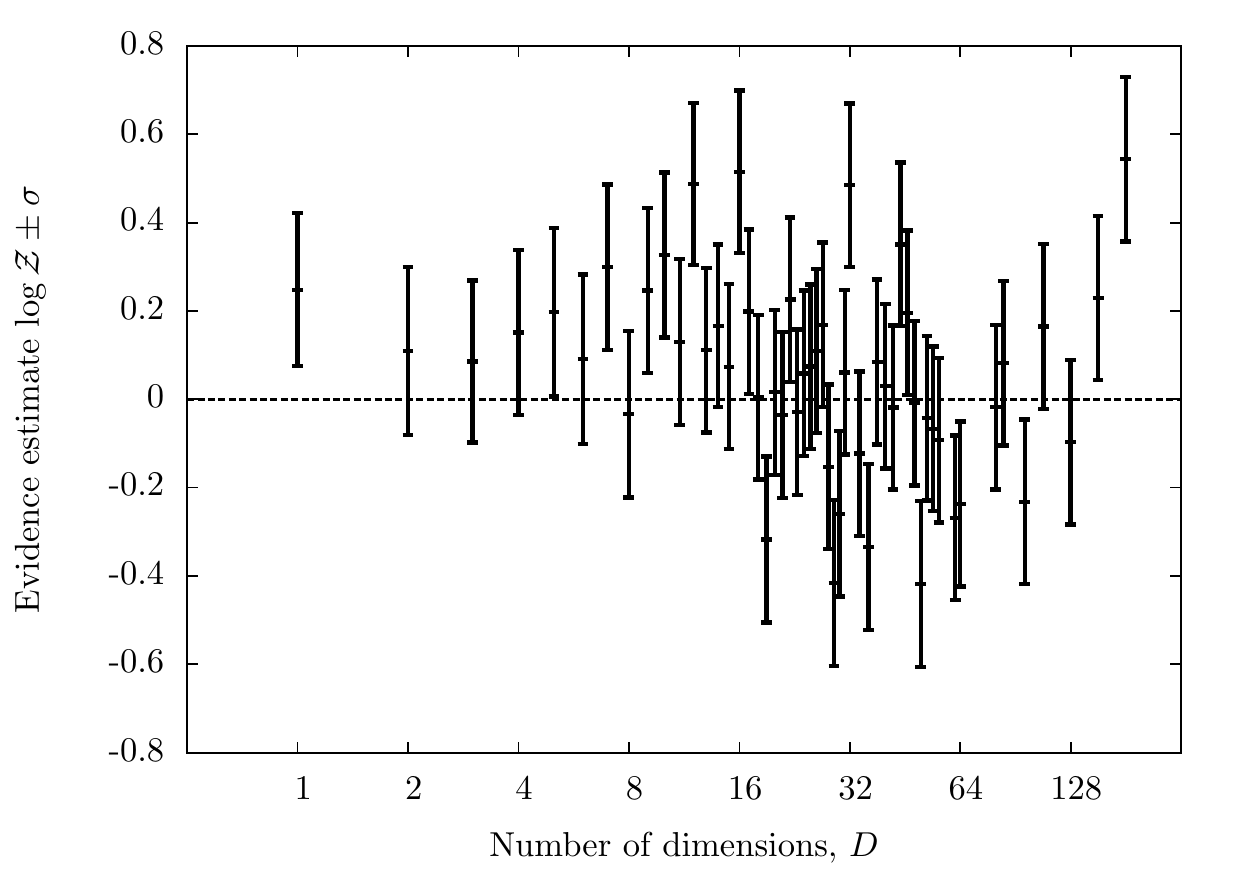}
  \caption{%
    Evidence estimates and errors produced by \PolyChord{} for a Gaussian likelihood as a function of dimensionality. The dashed line indicates the correct analytic evidence value.
    \label{fig:gaussian_evidences}
}
\end{figure}

\subsection{\CosmoChord}
\PolyChord{}'s real strength lies in its ability to exploit a
fast-slow hierarchy common in many cosmological applications. We have
successfully implemented \PolyChord{} within \CosmoMC{}, termed
\CosmoChord{}.  The traditional Metropolis-Hastings algorithm is
replaced with nested sampling. This implementation is available to
download from the link at the end of the paper.

This combination has been effectively implemented in multiple
cosmological applications in the latest Planck paper describing
constraints on inflation \citep{planck2015-a24}, including
application to a $37$-parameter reconstruction problem ($4$ slow, $19$
semi-slow, $14$ fast).

\section{Conclusions}
We have introduced \PolyChord{}, a novel nested sampling algorithm
tailored for high dimensional parameter spaces. It is able to fully
exploit a hierarchy of parameter speeds such as is found in \CosmoMC{}
and \CAMB{} \citep{CosmoMC,CAMB}. It utilises slice sampling at each
iteration to sample within the hard likelihood constraint of nested
sampling. It can identify and evolve separate modes of a posterior
semi-independently and is parallelised using \openMPI{}. We
demonstrate its efficacy on a toy problem. A further more detailed
paper will follow imminently.

\section*{Acknowledgements}
We would like to thank Farhan Feroz for numerous helpful discussions during the inception of the PolyChord algorithm.

\section*{Download Link}
\raggedright{PolyChord is available for download from:} \url{http://ccpforge.cse.rl.ac.uk/gf/project/polychord/} \\

\small
\bibliographystyle{apj}
\bibliography{polychord}

\label{lastpage}

\end{document}